\documentclass[twocolumn,preprintnumbers,amsmath,amssymb]{revtex4-1}

\usepackage{graphicx}
\usepackage{dcolumn}
\usepackage{bm}
\usepackage{rotating}

\usepackage{bbm}
\usepackage{verbatim}
\usepackage{afterpage}

\usepackage{color}
\definecolor{lightblue}{rgb}{0.2,0.2,0.7}
\definecolor{darkblue}{rgb}{0,0.25,0.5}
\definecolor{redbrown}{rgb}{0.875,0.25,0.125}
\definecolor{darkgreen}{rgb}{0,0.5,0}

\newcommand{\ket}[1]{\ensuremath{\vert #1  \rangle}}

\renewcommand{\b}[1]{\ensuremath{\mathbf{#1}}}

\begin{document}

\title{Quantum Monte Carlo calculations of electronic excitation energies: the case of the singlet $n \to \pi^*$ (CO) transition in acrolein}

\author{Julien Toulouse$^1$}\email{julien.toulouse@upmc.fr}
\author{Michel Caffarel$^2$}
\author{Peter Reinhardt$^1$}
\author{Philip E. Hoggan$^3$}
\author{C. J. Umrigar$^4$}
\affiliation{
$^1$Laboratoire de Chimie Th\'eorique, Universit\'e Pierre et Marie Curie and CNRS, Paris, France.\\
$^2$Laboratoire de Chimie et Physique Quantiques, IRSAMC, CNRS and Universit\'e de Toulouse, Toulouse, France.\\
$^3$LASMEA, CNRS and Universit\'e Blaise Pascal, Aubi\`ere, France.\\
$^4$Laboratory of Atomic and Solid State Physics, Cornell University, Ithaca, New York, USA.}

\date{\today}

\begin{abstract}
We report state-of-the-art quantum Monte Carlo calculations of the singlet $n \to \pi^*$ (CO) vertical excitation energy in the acrolein molecule, extending the recent study of Bouab\c{c}a {\it et al.} [J. Chem. Phys. {\bf 130}, 114107 (2009)]. We investigate the effect of using a Slater basis set instead of a Gaussian basis set, and of using state-average versus state-specific complete-active-space (CAS) wave functions, with or without reoptimization of the coefficients of the configuration state functions (CSFs) and of the orbitals in variational Monte Carlo (VMC).
It is found that, with the Slater basis set used here, both state-average and state-specific CAS(6,5) wave functions give an accurate excitation energy in diffusion Monte Carlo (DMC), with or without reoptimization of the CSF and orbital coefficients in the presence of the Jastrow factor. 
In contrast, the CAS(2,2) wave functions require reoptimization of the CSF and orbital coefficients to give a good DMC excitation energy.
Our best estimates of the vertical excitation energy are between 3.86 and 3.89 eV.
\end{abstract}

\maketitle

\section{Introduction}

Quantum Monte Carlo (QMC) methods (see, e.g., Refs.~\onlinecite{HamLesRey-BOOK-94,NigUmr-BOOK-99,FouMitNeeRaj-RMP-01}) constitute an alternative to standard quantum chemistry approaches for accurate calculations of the electronic structure of atoms, molecules and solids. The two most commonly used variants, variational Monte Carlo (VMC) and diffusion Monte Carlo (DMC), use a flexible trial wave function, generally consisting for atoms and molecules of a Jastrow factor multiplied by a short expansion in configuration state functions (CSFs), each consisting of a linear combination of Slater determinants. Although VMC and DMC have mostly been used for computing ground-state energies, excitation energies have been calculated as well (see, e.g., Refs~\onlinecite{GroRohMitLouCoh-PRL-01,AspElaGroLes-JCP-04,SchFil-JCP-04,SchBudFil-JCP-04,DruWilNeeGal-PRL-05,SceFil-PRB-06,CorJouIpaCasFilVel-JCP-07,TiaKenHooReb-JCP-08,TapTavRotFilCas-JCP-08,BouBenMayCaf-JCP-09,FilZacBud-JCTC-09,ZimTouZhaMusUmr-JCP-09}).

The simplest QMC calculations of excited states have been performed without reoptimizing the determinantal part of the wave function in the presence of the Jastrow factor. It has recently become possible to optimize in VMC both the Jastrow and determinantal parameters for excited states, either in a state-specific or a state-average approach~\cite{SchFil-JCP-04,SchBudFil-JCP-04,SceFil-PRB-06,CorJouIpaCasFilVel-JCP-07,TapTavRotFilCas-JCP-08,FilZacBud-JCTC-09,ZimTouZhaMusUmr-JCP-09}. Although this leads to very reliable excitation energies, reoptimization of the orbitals in VMC can be too costly for large systems.

In this context, Bouab\c{c}a {\it et al.}~\cite{BouBenMayCaf-JCP-09} studied how to obtain a reliable excitation energy in QMC for the singlet $n \to \pi^*$ (CO) vertical transition in the acrolein molecule without reoptimization of the determinantal part of the wave function. The acrolein molecule is the simplest member of the unsaturated aldehyde family whose photochemistry is of great interest. They showed that a good DMC excitation energy can be obtained by using {\it non-reoptimized} complete-active-space (CAS) wave functions if two conditions are fulfilled: a) The wave functions come from a {\it state-average} multiconfiguration self-consistent-field (MCSCF) calculation (using the same molecular orbitals for the two states is indeed expected to improve the compensation of errors due to the fixed-node approximation in the excitation energy); and b) a sufficiently large active space including all chemically relevant molecular orbitals for the excitation process is used. In comparison, all the small CAS wave functions and the large {\it state-specific} CAS wave functions (coming from two separate MCSCF calculations) were found to lead to quite unreliable DMC excitation energies, with a strong dependence on the size of the basis set. These results were obtained using standard all-electron QMC calculations with Gaussian basis sets, with orbitals appropriately modified near the nuclei to enforce the electron-nucleus cusp condition, in the same spirit as in Ref.~\onlinecite{ManLuc-JCP-01}.

In this work, we extend the study of Bouab\c{c}a {\it et al.} by testing the use of a Slater basis set and the effect of reoptimization of the determinantal part of the wave function in VMC. The use of Slater basis functions is motivated by the observation that they are capable of correctly reproducing the electron-nucleus cusp condition as well as having the correct exponential decay at large distances. In contrast, Gaussian basis functions have no cusp at the nucleus and a too rapid decay at large distances. As regards the effect of reoptimization, conclusions about the validity of using non-reoptimized CAS wave functions are drawn. The paper is organized as follows. In Section~\ref{sec:methodology}, we explain the methodology used. In Section~\ref{sec:results}, we present and discuss our results. Finally, Section~\ref{sec:conclusion} summarizes our conclusions.

\section{Methodology}
\label{sec:methodology}

\begin{figure}[t]
\includegraphics[scale=0.35,angle=0]{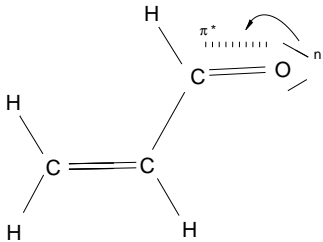}
\caption{Schematic representation of the singlet $n \to \pi^*$ excitation in the CO moiety of the acrolein molecule.}
\label{fig:c3h4o}
\end{figure}

We are concerned with the vertical electronic transition in the acrolein (or propenal) molecule, CH$_2$=CH-CHO (symmetry group $C_s$), from the spin-singlet ground state (symmetry A') to the first spin-singlet excited state (A"). This transition is identified as the excitation of an electron from the lone pair ($n$) of the oxygen to the antibonding $\pi^*$ orbital of the CO moiety. We use the {\it s-trans} experimental geometry of Ref.~\onlinecite{BloGraBau-JACS-84}, obtained by microwave spectroscopy in the gas phase.

We use Jastrow-Slater wave functions parametrized as~\cite{TouUmr-JCP-07,TouUmr-JCP-08}
\begin{equation}
\ket{\Psi(\b{p})} = \hat{J}(\bm{\alpha}) e^{\hat{\kappa} (\bm{\kappa})} \sum_{I=1}^{N_{\rm CSF}} c_I \ket{C_I},
\end{equation}
where $\hat{J}(\bm{\alpha})$ is a Jastrow factor operator, $e^{\hat{\kappa}(\bm{\kappa})}$ is the orbital rotation operator and $\ket{C_I}$ are CSFs. Each CSF is a symmetry-adapted linear combination of Slater determinants of single-particle orbitals which are expanded in Slater basis functions. The parameters $\b{p}=(\bm{\alpha},\b{c},\bm{\kappa})$ that are optimized are the Jastrow parameters $\bm{\alpha}$, the CSF coefficients $\b{c}$ and the orbital rotation parameters $\bm{\kappa}$. The exponents of the basis functions are kept fixed in this work. We use a Jastrow factor consisting of the exponential of the sum of electron-nucleus, electron-electron, and electron-electron-nucleus terms, written as systematic polynomial and Pad\'e expansions~\cite{Umr-UNP-XX} (see also Refs.~\onlinecite{FilUmr-JCP-96} and~\onlinecite{GucSanUmrJai-PRB-05}).

For each state, we start by generating standard restricted Hartree-Fock (RHF), and state-average and state-specific MCSCF wave functions with a complete active space generated by distributing $N$ valence electrons in $M$ valence orbitals [CAS($N$,$M$)], using the quantum chemistry program GAMESS~\cite{SchBalBoaElbGorJenKosMatNguSuWinDupMon-JCC-93}. As in Ref.~\onlinecite{BouBenMayCaf-JCP-09}, we consider a minimal CAS(2,2) active space containing the two molecular orbitals $n$ (A') and $\pi_\text{CO}^*$ (A") involved in the excitation, and a larger CAS(6,5) active space containing the 5 molecular orbitals that are expected to be chemically relevant: $\pi_\text{CO}$ (A"), $n$ (A'), $\pi_\text{CC}$ (A"), $\pi_\text{CO}^*$ (A"), $\pi_\text{CC}^*$ (A"). Note that, since the two states have different symmetries, the purpose behind using the state-average procedure is not the usual one of avoiding a variational collapse of the excited state onto the ground state, but rather to possibly improve the compensation of errors in the excitation energy by using the same molecular orbitals for the two states. We use the triple-zeta quality VB1 Slater basis of Ema {\it et al.}~\cite{EmaGarRamLopFerMeiPal-JCC-03}. For C and O, this basis contains two $1s$, three $2s$, three $2p$ and one $3d$ sets of functions; for H, it contains three $1s$ and one $2p$ sets of functions. Each Slater function is actually approximated by a fit to 10 Gaussian functions~\cite{HehStePop-JCP-69,Ste-JCP-70,KolReiAss-JJJ-XX} in GAMESS. Theses wave functions are then multiplied by the Jastrow factor, imposing the electron-electron cusp condition, and QMC calculations are performed with the program CHAMP~\cite{Cha-PROG-XX} using the true Slater basis set rather than its Gaussian expansion. The wave function parameters are optimized with the linear energy minimization method in VMC~\cite{TouUmr-JCP-07,UmrTouFilSorHen-PRL-07,TouUmr-JCP-08}, using an accelerated Metropolis algorithm~\cite{Umr-PRL-93,Umr-INC-99}. Two levels of optimization are tested: optimization of only the Jastrow factor while keeping the CSF and orbital parameters at their RHF or MCSCF values, and simultaneous optimization of the Jastrow, CSF and orbital parameters. For all wave functions, even the state-average ones, we always optimize a separate Jastrow factor for each state, rather than a common Jastrow factor for the two states. Although the electron-nucleus cusp condition is not enforced during the optimization in our current implementation, the orbitals obtained from Slater basis functions usually nearly satisfy the cusp condition. Once the trial wave functions have been optimized, we perform DMC calculations within the short-time and fixed-node (FN) approximations (see, e.g., Refs.~\onlinecite{GriSto-JCP-71,And-JCP-75,And-JCP-76,ReyCepAldLes-JCP-82,MosSchLeeKal-JCP-82}). We use an efficient DMC algorithm with very small time-step errors~\cite{UmrNigRun-JCP-93}. For a given trial wave function, the evolution of the ground- and excited-state total DMC energies and of the corresponding excitation energy when the imaginary time step $\tau$ is decreased from 0.01 to 0.001 hartree$^{-1}$ is shown in Table~\ref{tab:timestep}. While the time-step bias is clearly seen for the total energies, it largely cancels out for the excitation energy for all the time steps tested here and cannot be resolved within the statistical uncertainty. In the following, we always use an imaginary time step of $\tau=0.001$ hartree$^{-1}$. Note that the Jastrow factor does not change the nodes of the wave function, and therefore it has no direct effect on the fixed-node DMC total energy (aside from of course the time-step bias and the population-control bias). Improving the trial wave function by optimization of the Jastrow factor is nevertheless important for DMC calculations in order to reduce the fluctuations and to make the time-step error very small and the population-control bias negligible. Of course, when the Jastrow factor is optimized together with the CSF and/or orbital parameters, then it has an indirect effect through those parameters on the nodes of the wave function.

\begin{table*}[t]
\caption{Ground-state energy $E_0$, first excited-state energy $E_1$, and vertical excitation energy $E_1 - E_0$ for the singlet $n\to\pi^*$ transition in the acrolein molecule at the experimental geometry calculated in DMC with different time steps $\tau$ using the VB1 Slater basis set and a state-specific Jastrow-Slater CAS(6,5) wave function with Jastrow, CSF and orbital parameters optimized by energy minimization in VMC.}
\label{tab:timestep}
\begin{tabular}{lccc}
\hline
\hline
$\tau$ (hartree$^{-1}$)                        &  $E_0$ (hartree) &       $E_1$ (hartree)   &   $E_1 - E_0$ (eV) \\
\hline

0.01                                           & -191.8734(4)      &       -191.7312(4)      &      3.87(2)                \\
0.005                                          & -191.8753(4)      &       -191.7319(4)      &      3.90(2)                \\
0.0025                                         & -191.8762(4)      &       -191.7330(4)      &      3.90(2)                \\
0.001                                          & -191.8769(3)      &       -191.7350(3)      &      3.86(1)                \\
\hline
\hline
\end{tabular}
\end{table*}

\section{Results and discussion}
\label{sec:results}

Table~\ref{tab:energies} reports the ground-state energy $E_0$, the first excited-state energy $E_1$, and the excitation energy $E_1-E_0$ calculated by different methods. Since the excited state is an spin-singlet open-shell state, it cannot be described by a restricted single-determinant wave function; however, we report single-determinant results for the ground state for comparison of total energies. We take our best estimates of the vertical excitation energy to be those obtained with the CAS(6,5) wave functions in DMC. They range from 3.86 to 3.89 eV, depending whether a state-average or state-specific approach is used and whether the determinantal part of the wave function is reoptimized in QMC. Previously reported calculations include (a) time-dependent density-functional theory (TDDFT): 3.66 eV~\cite{AquBarRoo-JCP-03} and 3.78 eV~\cite{AidMogNilJohMikChrSodKon-JCP-08}; (b) complete-active-space second-order perturbation theory (CASPT2): 3.63 eV~\cite{AquBarRoo-JCP-03}, 3.69 eV~\cite{MarLosFdeAgu-JCP-04}, and 3.77 eV~\cite{LosFdeAguMar-JPC-07}; (c) multireference configuration interaction: 3.85 eV~\cite{DomMulDalLisDieKla-TCA-04}; (d) different variants of coupled cluster: 3.83 eV~\cite{SahEhaNak-JCP-06}, 3.93 eV~\cite{AidMogNilJohMikChrSodKon-JCP-08}, 3.75 eV~\cite{AidMogNilJohMikChrSodKon-JCP-08}. The most recent experimental estimate is 3.69 eV, which corresponds to the maximum in the UV absorption band in gas phase and which is in agreement with previous experimental data~\cite{BlaYouRoo-JACS-37,Inu-BCSJ-60,Hol-SA-63,MosYabBon-TEC-66}. Beside different treatment of electron correlation, the discrepancies between these values may be due to the high sensitivity of the excitation energy to the C=C and C=O bond lengths~\cite{AidMogNilJohMikChrSodKon-JCP-08}. Moreover, the comparison with experiment relies on the approximation that the vertical excitation energy corresponds to the maximum of the broad UV absorption band. In view of all these data, a safe estimate range for the exact vertical excitation energy is from about 3.60 to 3.90 eV.

Even without reoptimization of the CSF and orbital coefficients, our state-specific Jastrow-Slater CAS(6,5) wave functions give a DMC excitation energy, 3.88(2) eV, as accurate as the one obtained with the fully optimized wave functions, even though the total energies $E_0$ and $E_1$ are about 20 mhartree higher. Also, our non-reoptimized state-average Jastrow-Slater CAS(6,5) wave functions give an essentially identical DMC excitation energy of 3.89(2) eV. This agrees well with the DMC result of Bouab\c{c}a {\it et al.}~\cite{BouBenMayCaf-JCP-09}, 3.86(7) eV, obtained with non-reoptimized state-average Jastrow-Slater CAS(6,5) wave functions with a Gaussian basis set.

\begin{table*}[t]
\caption{Ground-state energy $E_0$, first excited-state energy $E_1$, and vertical excitation energy $E_1 - E_0$ for the singlet $n\to\pi^*$ transition in the acrolein molecule at the experimental geometry calculated by different methods using the VB1 Slater basis set. The QMC calculations are done with Jastrow-Slater wave functions using a single determinant (JSD), or a state-average (SA) or state-specific (SS) complete-active-space multideterminant expansion (JCAS). The lists of parameters optimized by energy minimization in VMC are indicated within square brackets: Jastrow (J), CSF coefficients (c), and orbitals (o). For comparison, the DMC results of Ref.~\onlinecite{BouBenMayCaf-JCP-09} obtained with state-average CAS(6,5) wave functions and a Gaussian basis set are also shown.}
\label{tab:energies}
\begin{tabular}{lccc}
\hline
\hline
                           &   $E_0$ (hartree)    &       $E_1$ (hartree)      &   $E_1 - E_0$ (eV) \\
\hline
RHF                        & -190.83430261     &                         &                             \\
MCSCF CAS(2,2) SA          & -190.82258836     &      -190.68568203      &      3.73                   \\
MCSCF CAS(2,2) SS          & -190.83891553     &      -190.71709289      &      3.31                   \\
MCSCF CAS(6,5) SA          & -190.88736483     &      -190.74691372      &      3.82                   \\
MCSCF CAS(6,5) SS          & -190.89520291     &      -190.75181511      &      3.90                   \\

\\
VMC JSD    [J]             & -191.7107(5)      &                         &                             \\
VMC JSD    [J+o]           & -191.7636(5)      &                         &                             \\
VMC JCAS(2,2) SA [J]       & -191.7121(5)      &       -191.5619(5)      &      4.09(2)                \\
VMC JCAS(2,2) SS [J]       & -191.7099(5)      &       -191.5652(5)      &      3.94(2)                \\
VMC JCAS(2,2) SS [J+c+o]   & -191.7643(5)      &       -191.6247(5)      &      3.80(2)                \\
VMC JCAS(6,5) SA [J]       & -191.7182(5)      &       -191.5747(5)      &      3.90(2)                \\
VMC JCAS(6,5) SS [J]       & -191.7221(5)      &       -191.5776(5)      &      3.93(2)                \\
VMC JCAS(6,5) SS [J+c+o]   & -191.7795(5)      &       -191.6342(5)      &      3.95(2)                \\
\\
DMC JSD    [J]             & -191.8613(4)      &                                                       \\
DMC JSD    [J+o]           & -191.8698(3)      &                                                       \\
DMC JCAS(2,2) SA [J]       & -191.8608(5)      &       -191.7133(5)      &      4.01(2)                \\
DMC JCAS(2,2) SS [J]       & -191.8606(4)      &       -191.7113(4)      &      4.06(2)                \\
DMC JCAS(2,2) SS [J+c+o]   & -191.8700(3)      &       -191.7293(3)      &      3.83(1)                \\
DMC JCAS(6,5) SA [J]       & -191.8568(5)      &       -191.7138(5)      &      3.89(2)                \\
DMC JCAS(6,5) SS [J]       & -191.8585(4)      &       -191.7160(4)      &      3.88(2)                \\
DMC JCAS(6,5) SS [J+c+o]   & -191.8769(3)      &       -191.7350(3)      &      3.86(1)                \\
\\
DMC JCAS(6,5) SA [J]$^a$   & -191.8504(20)     &       -191.7086(23)     &      3.86(7)                \\
\\
Experimental estimate$^b$  &                   &                         &      3.69                   \\
\hline
\hline
$^a$QMC calculations with a Gaussian basis, Ref.~\onlinecite{BouBenMayCaf-JCP-09}.\\
$^b$Maximum in the UV absorption band in gas phase, Ref.~\onlinecite{AidMogNilJohMikChrSodKon-JCP-08}.
\end{tabular}
\end{table*}

Thus it appears possible to obtain an accurate excitation energy using non-reoptimized state-specific CAS(6,5) wave functions in DMC. This is different from what was observed in Ref.~\onlinecite{BouBenMayCaf-JCP-09} where state-specific CAS(6,5) wave functions were found to give unreliable excitation energies. The difference is that we use here a Slater basis set rather than the Gaussian basis set employed in Ref.~\onlinecite{BouBenMayCaf-JCP-09}. Even though the Gaussian basis contains more basis functions than the VB1 Slater basis, it gives a higher DMC energy for both states and tends to favor one state over the other in state-specific calculations. This example shows the importance of using a well-balanced basis set in state-specific calculations, even in DMC.

We comment now on the results obtained with the CAS(2,2) wave functions. The state-specific MCSCF CAS(2,2) excitation energy, 3.31 eV, is a strong underestimate. The corresponding VMC and DMC state-specific calculations without reoptimization of the CSF and orbital coefficients, give slightly overestimated excitation energies, 3.94(2) and 4.06(2) eV, respectively. Whereas the state-average MCSCF CAS(2,2) calculation gives a much better excitation energy, 3.73 eV, compared to the state-specific MCSCF calculation, the non-reoptimized state-average CAS(2,2) wave functions do not seem to improve the excitation energies in VMC and DMC. In fact, they give a worse VMC excitation energy of 4.09(2) eV, and a DMC excitation energy of 4.01(2) eV which is not significantly better than with the non-reoptimized state-specific wave functions.

The excitation energies obtained from the CAS(6,5) wave functions depend very little on whether a) they are calculated in MCSCF, VMC or DMC,
b) the state-average or the state-specific approach is employed, and
c) the CSF and orbital coefficients are reoptimized or not in the presence of the Jastrow factor.
In contrast, the excitation energies obtained from CAS(2,2) wave functions do depend on all of the above and, in particular the reoptimization of the CSF and orbital coefficients in the presence of the Jastrow factor significantly improves the VMC and DMC excitation energies, to 3.80(2) and 3.83(1) eV, respectively. The importance of reoptimizing in VMC the CAS(2,2) expansions but not the CAS(6,5) expansions suggests that the Jastrow factor includes important correlation effects that are present in CAS(6,5) but not in CAS(2,2).

Finally, we note that without reoptimization of the determinantal part of the wave functions, the ground-state VMC and DMC energies can actually increase when going from a single-determinant wave function to a CAS(2,2) or CAS(6,5) wave function. This behavior has been observed in other systems as well, e.g. in C$_2$ and Si$_2$~\cite{UmrTouFilSorHen-PRL-07}. Of course, if the CSF and orbital coefficients are reoptimized in VMC, then the VMC total energies must decrease monotonically upon increasing the number of CSFs. In practice, it is found that the DMC also decrease monotonically although there is in principle no guarantee that optimization in VMC necessarily improves the nodes of the wave function.

\section{Conclusion}
\label{sec:conclusion}

In this work, we have extended the study of Bouab\c{c}a {\it et al.}~\cite{BouBenMayCaf-JCP-09} on how to obtain a reliable excitation energy in QMC for the singlet $n \to \pi^*$ (CO) vertical transition in the acrolein molecule. We have tested the use of a Slater basis set and the effect of reoptimization of the determinantal part of the wave function in VMC and of the corresponding changes in the nodal structure in fixed-node DMC. Putting together the conclusions of the study of Bouab\c{c}a {\it et al.} and the present one, we can summarize the findings on acrolein as follows:\\
a) It is possible to obtain an accurate DMC excitation energy with non-reoptimized CAS wave functions, provided that a sufficiently large chemically relevant active
space is used. In the case of a too small active space, reoptimization of the CSF and orbital coefficients in the presence of the Jastrow factor appears to be necessary in order to get a good DMC excitation energy.\\
b) When using Gaussian basis sets of low or intermediate quality, reliable DMC excitation energies could be obtained only by using state-average wave functions (i.e., with the same molecular orbitals for the two states). In contrast, when using a good quality Slater basis set such as the VB1 basis, state-specific wave functions were found to also give reliable DMC excitation energies. Thus, this provides some support for using Slater, rather than Gaussian, basis sets in all-electron QMC calculations. Note that other authors also advocate the use of Slater basis sets in all-electron QMC calculations (see, e.g., Refs.~\onlinecite{GalHanCohCha-CPL-05,GalHanLes-MP-06,NemTowNee-JCP-10}).\\ 

It remains to check whether these conclusions are generally true for other systems. It would be indeed desirable for calculations on large molecular systems if accurate DMC excitation energies could be obtained with state-specific or state-average CAS expansions without the need of an expensive reoptimization of the determinantal part of the wave functions in QMC.

\section*{Acknowledgments}
Most QMC calculations have been done on the IBM Blue Gene of Forschungszentrum J\"ulich (Germany) within the DEISA project STOP-Qalm.
CJU acknowledges support from NSF grant number CHE-1004603.


\bibliographystyle{apsrev4-1}

\end{document}